

\def\singlespace{\normalbaselines}
\def\oneandahalfspace{\baselineskip=1.15\normalbaselineskip plus 1pt
\lineskip=2pt\lineskiplimit=1pt}

\def\np{\vfill\eject}
\def\nl{\hfil\break}

\def\nofirstpagenoten{\nopagenumbers\footline={\ifnum\pageno>1\tenrm
\hss\folio\hss\fi}}
\def\nofirstpagenotwelve{\nopagenumbers\footline={\ifnum\pageno>1\twelverm
\hss\folio\hss\fi}}
\def\leaderfill{\leaders\hbox to 1em{\hss.\hss}\hfill}
\def\ft#1#2{{\textstyle{{#1}\over{#2}}}}
\def\frac#1/#2{\leavevmode\kern.1em
\raise.5ex\hbox{\the\scriptfont0 #1}\kern-.1em/\kern-.15em
\lower.25ex\hbox{\the\scriptfont0 #2}}
\def\sfrac#1/#2{\leavevmode\kern.1em
\raise.5ex\hbox{\the\scriptscriptfont0 #1}\kern-.1em/\kern-.15em
\lower.25ex\hbox{\the\scriptscriptfont0 #2}}


\parindent=20pt
\def\narrow{\advance\leftskip by 40pt \advance\rightskip by 40pt}

\def\AB{\bigskip
        \centerline{\bf ABSTRACT}\medskip\narrow}
\def\nonarrower{\advance\leftskip by -40pt\advance\rightskip by -40pt}
\def\AE{\bigskip\nonarrower}

\def\boxit#1{\vbox{\hrule\hbox{\vrule\kern3pt
        \vbox{\kern3pt#1\kern3pt}\kern3pt\vrule}\hrule}}

\def\gtorder{\mathrel{\raise.3ex\hbox{$>$}\mkern-14mu
             \lower0.6ex\hbox{$\sim$}}}
\def\ltorder{\mathrel{\raise.3ex\hbox{$<$}|mkern-14mu
             \lower0.6ex\hbox{\sim$}}}
\def\dalemb#1#2{{\vbox{\hrule height .#2pt
        \hbox{\vrule width.#2pt height#1pt \kern#1pt
                \vrule width.#2pt}
        \hrule height.#2pt}}}

\font\fourteentt=cmtt10 scaled \magstep2
\font\fourteenbf=cmbx12 scaled \magstep1
\font\fourteenrm=cmr12 scaled \magstep1
\font\fourteeni=cmmi12 scaled \magstep1
\font\fourteenss=cmss12 scaled \magstep1
\font\fourteensy=cmsy10 scaled \magstep2
\font\fourteensl=cmsl12 scaled \magstep1
\font\fourteenex=cmex10 scaled \magstep2
\font\fourteenit=cmti12 scaled \magstep1
\font\twelvett=cmtt10 scaled \magstep1 \font\twelvebf=cmbx12
\font\twelverm=cmr12 \font\twelvei=cmmi12
\font\twelvess=cmss12 \font\twelvesy=cmsy10 scaled \magstep1
\font\twelvesl=cmsl12 \font\twelveex=cmex10 scaled \magstep1
\font\twelveit=cmti12
\font\tenss=cmss10
 
 \font\ninebf=cmbx7 scaled \magstep1
\font\ninerm=cmr7 scaled \magstep1 \font\ninei=cmmi7 scaled \magstep1
\font\ninesy=cmsy7 scaled \magstep1 
\font\eightrm=cmr7 scaled 1140 
 
\font\sevenbf=cmbx7 \font\sevenrm=cmr7 \font\seveni=cmmi7
\font\sevensy=cmsy7 

\catcode`@=11
\newskip\ttglue
\newfam\ssfam

\def\fourteenpoint{\def\rm{\fam0\fourteenrm}
\textfont0=\fourteenrm \scriptfont0=\tenrm \scriptscriptfont0=\sevenrm
\textfont1=\fourteeni \scriptfont1=\teni \scriptscriptfont1=\seveni
\textfont2=\fourteensy \scriptfont2=\tensy \scriptscriptfont2=\sevensy
\textfont3=\fourteenex \scriptfont3=\fourteenex \scriptscriptfont3=\fourteenex
\def\it{\fam\itfam\fourteenit} \textfont\itfam=\fourteenit
\def\sl{\fam\slfam\fourteensl} \textfont\slfam=\fourteensl
\def\bf{\fam\bffam\fourteenbf} \textfont\bffam=\fourteenbf
\scriptfont\bffam=\tenbf \scriptscriptfont\bffam=\sevenbf
\def\tt{\fam\ttfam\fourteentt} \textfont\ttfam=\fourteentt
\def\ss{\fam\ssfam\fourteenss} \textfont\ssfam=\fourteenss
\tt \ttglue=.5em plus .25em minus .15em
\normalbaselineskip=16pt
\abovedisplayskip=16pt plus 4pt minus 12pt
\belowdisplayskip=16pt plus 4pt minus 12pt
\abovedisplayshortskip=0pt plus 4pt
\belowdisplayshortskip=9pt plus 4pt minus 6pt
\parskip=5pt plus 1.5pt
\setbox\strutbox=\hbox{\vrule height12pt depth5pt width0pt}
\let\sc=\tenrm
\let\big=\fourteenbig \normalbaselines\rm}
\def\fourteenbig#1{{\hbox{$\left#1\vbox to12pt{}\right.\n@space$}}}

\def\twelvepoint{\def\rm{\fam0\twelverm}
\textfont0=\twelverm \scriptfont0=\ninerm \scriptscriptfont0=\sevenrm
\textfont1=\twelvei \scriptfont1=\ninei \scriptscriptfont1=\seveni
\textfont2=\twelvesy \scriptfont2=\ninesy \scriptscriptfont2=\sevensy
\textfont3=\twelveex \scriptfont3=\twelveex \scriptscriptfont3=\twelveex
\def\it{\fam\itfam\twelveit} \textfont\itfam=\twelveit
\def\sl{\fam\slfam\twelvesl} \textfont\slfam=\twelvesl
\def\bf{\fam\bffam\twelvebf} \textfont\bffam=\twelvebf
\scriptfont\bffam=\ninebf \scriptscriptfont\bffam=\sevenbf
\def\tt{\fam\ttfam\twelvett} \textfont\ttfam=\twelvett
\def\ss{\fam\ssfam\twelvess} \textfont\ssfam=\twelvess
\tt \ttglue=.5em plus .25em minus .15em
\normalbaselineskip=14pt
\abovedisplayskip=14pt plus 3pt minus 10pt
\belowdisplayskip=14pt plus 3pt minus 10pt
\abovedisplayshortskip=0pt plus 3pt
\belowdisplayshortskip=8pt plus 3pt minus 5pt
\parskip=3pt plus 1.5pt
\setbox\strutbox=\hbox{\vrule height10pt depth4pt width0pt}
\let\sc=\ninerm
\let\big=\twelvebig \normalbaselines\rm}
\def\twelvebig#1{{\hbox{$\left#1\vbox to10pt{}\right.\n@space$}}}

\def\tenpoint{\def\rm{\fam0\tenrm}
\textfont0=\tenrm \scriptfont0=\sevenrm \scriptscriptfont0=\fiverm
\textfont1=\teni \scriptfont1=\seveni \scriptscriptfont1=\fivei
\textfont2=\tensy \scriptfont2=\sevensy \scriptscriptfont2=\fivesy
\textfont3=\tenex \scriptfont3=\tenex \scriptscriptfont3=\tenex
\def\it{\fam\itfam\tenit} \textfont\itfam=\tenit
\def\sl{\fam\slfam\tensl} \textfont\slfam=\tensl
\def\bf{\fam\bffam\tenbf} \textfont\bffam=\tenbf
\scriptfont\bffam=\sevenbf \scriptscriptfont\bffam=\fivebf
\def\tt{\fam\ttfam\tentt} \textfont\ttfam=\tentt
\def\ss{\fam\ssfam\tenss} \textfont\ssfam=\tenss
\tt \ttglue=.5em plus .25em minus .15em
\normalbaselineskip=12pt
\abovedisplayskip=12pt plus 3pt minus 9pt
\belowdisplayskip=12pt plus 3pt minus 9pt
\abovedisplayshortskip=0pt plus 3pt
\belowdisplayshortskip=7pt plus 3pt minus 4pt
\parskip=0.0pt plus 1.0pt
\setbox\strutbox=\hbox{\vrule height8.5pt depth3.5pt width0pt}
\let\sc=\eightrm
\let\big=\tenbig \normalbaselines\rm}
\def\tenbig#1{{\hbox{$\left#1\vbox to8.5pt{}\right.\n@space$}}}
\let\rawfootnote=\footnote \def\footnote#1#2{{\rm\parskip=0pt\rawfootnote{#1}
{#2\hfill\vrule height 0pt depth 6pt width 0pt}}}

\def\tenfoot{\tenpoint\hskip-\parindent\hskip-.1cm}

\overfullrule=0pt
\twelvepoint
\def\sbullet{\raise.2em\hbox{$\scriptscriptstyle\bullet$}}
\nofirstpagenotwelve
\hsize=16.5 truecm
\baselineskip 15pt

\def\ft#1#2{{\textstyle{{#1}\over{#2}}}}

\def\ket#1{\big| #1\big\rangle}

\def\phys{\big|{\rm phys}\big\rangle}

\def\del{\partial}

\def\phys{\big|\hbox{phys}\big\rangle}

\oneandahalfspace
\rightline{CTP TAMU--43/93}
\rightline{KUL--TF--93/34}
\rightline{hep-th/9308114}
\rightline{August 1993}

\vskip 2truecm
\centerline{\bf Higher-spin strings and $W$ minimal models}
\vskip 1.5truecm
\centerline{H. Lu, C.N. Pope,\footnote{$^*$}{\tenfoot Supported in part
by the U.S. Department of Energy, under
grant DE-FG05-91ER40633.} X.J. Wang\footnote{}{\tenfoot }}
\vskip 1.5truecm
\centerline{\it Center
for Theoretical Physics,
Texas A\&M University,}
\centerline{\it College Station, TX 77843--4242, USA}
\vskip 2truecm
\centerline{K. Thielemans}
\vskip 1.5truecm
\centerline{\it Instituut voor Theoretische Fysica,}
\centerline{\it Celestijnenlaan 200D, B-3001 Leuven, Belgium.}

\vskip 1.5truecm
\AB\singlespace
   We study the spectrum of physical states for higher-spin generalisations of
string theory, based on two-dimensional theories with local spin-2 and
spin-$s$ symmetries.  We explore the relation of the resulting effective
Virasoro string theories to certain $W$ minimal models.  In particular, we
show how the highest-weight states of the $W$ minimal models decompose into
Virasoro primaries.
\AE\oneandahalfspace

\np
\noindent
{\bf 1. Introduction}
\bigskip

     In a recent paper it was shown that the BRST operator $Q_B$ for the $W_3$
algebra, in the case where the matter currents are realised in terms of scalar
fields,  admits a considerable simplification if one performs a redefinition
that mixes the ghost fields and one of the matter fields, called $\varphi$,
that plays a special r\^ole in the realisation [1].  In particular, the BRST
operator written in terms of the redfined fields has a double grading; {\it
i.e.}\ $Q_B=Q_0 + Q_1$, where $Q_0$ has grade $(1,0)$ and $Q_1$ has grade
$(0,1)$, with $(p,q)$ denoting the grading of an operator with ghost number $p$
for the spin-2 ghost system and ghost number $q$ for the spin-3 ghost system.

     The form of the BRST operator for the $W_3$ algebra immediately suggests a
generalisation to the case where the two matter currents have spins 2 and $s$
rather than 2 and 3 [2].  In [2], such BRST  operators were explicitly
constructed,  for the cases $s=4$, 5 and 6, and some aspects of the physical
spectra of the associated generalisations of string theory were explored.
Presumably the construction would work for higher values of $s$ too.  It should
be emphasised that such BRST operators can apparently be constructed for
general $s$, even though closed $W$ algebras of the form $W_{(2,s)}$ do not
exist for general values of $s$.  Evidently the conditions for the existence
of a nilpotent BRST operator are weaker than the conditions for the existence
of a closed operator algebra. (A previous example of this kind of phenomenon,
where a BRST operator exists even when the matter system does not generate a
closed algebra at the quantum level, was found in the context of the
``non-critical'' $W_3$ string discussed in [3,4,5].)

     It was shown in [2] that for $s=3$, 4, 5 and 6, there is a BRST operator
for a spin-2 plus spin-$s$ system that gives rise to a higher-spin string
theory
that appears to be unitary. It was conjectured that this theory admits  an
interpretation as an effective Virasoro string theory, with central charge
$c^{\rm eff}=26 - {2(s-2)\over (s+1)}$, coupled to a certain minimal model with
central charge ${2(s-2)\over (s+1)}$ realised by the special matter field
$\varphi$ and the spin-$s$ ghost system.  Some arguments were presented in [2]
that suggested that the minimal model in  question is the lowest unitary
$W_{s-1}$ minimal model ({\it i.e.}\ the $W_{s-1}$ generalisation of the Ising
model).  It is expected that the results should generalise to all $s$.

     In this paper, we develop the idea further, and study the physical states
in detail for $s=4$, 5 and 6.  In particular, for the spin-2 plus spin-4
string, we find further evidence that it is associated with the lowest $W_3$
minimal model, which has central charge $c=\ft45$.  We do this by finding a
primary spin-3 operator in the physical spectrum which, together with the
energy-momentum tensor, generates the $W_3$ algebra at $c=\ft45$.  We then
check that all the low-lying physical states (up to level 9) are either
highest-weight states of this $W_3$ minimal model with  $c=\ft45$, or else
they are $W$ descendants.  This example is somewhat non-generic in having
$c<1$, so the $W$ fields decompose into a {\it finite} number of Virasoro
primaries, namely into a subset of those of the 3-state Potts model.
Nonetheless, it provides a nice illustration of the manner in which the
highest-weight states of the $W$ model decompose into a larger set of Virasoro
primaries, the extra Virasoro primaries corresponding to operators that can be
written as descendants under the $W$ currents.  Similarly, for the spin-2 plus
spin-5 string, and the spin-2 plus spin-6 string, we show that their physical
spectra include primary operators that generate the $W_4$ algebra at $c=1$,
and the $W_5$ algebra at $c=\ft87$ respectively.  Again, we check that the
low-lying physical states are either highest weight under the currents of the
$W$ algebra, or else $W$ descendants.

     We have made extensive use of the Mathematica package OPEdefs [6] for
the conformal field theoretic calculations in this paper.

\bigskip
\noindent{\bf 2. Spin-2 plus spin-$s$ Strings}
\medskip
\noindent{\it 2.1 Generalities}
\bigskip

      We begin by introducing the usual $(b,c)$ ghost system for the spin-2
current, and the $(\beta,\gamma)$ ghost system for the spin-$s$ current.  Note
that $\beta$ therefore has spin $s$, and $\gamma$ has spin $(1-s)$.  The BRST
operator for the spin-2 plus spin-$s$ string then takes the form [2]:
$$
\eqalignno{ Q_B&= Q_0 + Q_1,&(2.1)\cr
Q_0&=\oint dz\, c \Big(T^{\rm eff} +T_{\varphi} + T_{\gamma,\beta} + \ft12
T_{c,b} \Big), &(2.2)\cr
Q_1&=\oint dz\, \gamma \, F(\varphi,\beta,\gamma),&(2.3)\cr}
$$
where the energy-momentum tensors are given by
$$
\eqalignno{
T_\varphi&\equiv -\ft12 (\del\varphi)^2 -\alpha\, \del^2\varphi, &(2.4)\cr
T_{\gamma,\beta}&\equiv -s\, \beta\,\del\gamma -(s-1)\, \del\beta\, \gamma,
&(2.5)\cr
T_{c,b}&\equiv -2\, b\, \del c - \del b\, c, &(2.6)\cr
T^{\rm eff} &\equiv -\ft12 \del X^\mu\, \del X^\nu\, \eta_{\mu\nu} -
i a_\mu\, \del^2 X^\mu. &(2.7)\cr}
$$
The operator $F(\varphi,\beta,\gamma)$ has spin $s$ and ghost number zero.
Because of the grading discussed in the introduction, it follows that one will
have $Q_0^2=Q_1^2=\{Q_0,Q_1\}=0$.  The first of these conditions is satisfied
provided that the total central charge vanishes, {\it i.e.}\
$$
0=-26 -2(6s^2-6s+1) + 1+12\alpha^2 + c^{\rm eff}.\eqno(2.8)
$$
The remaining two nilpotency conditions determine the precise form of the
operator $F(\varphi,\beta,\gamma)$ appearing in (2.3).  Solutions for $s=4$, 5
and 6 were found in [2].  (In fact in [2] it was found that there are two
different nilpotent BRST operators when $s=4$, and four different ones when
$s=6$.  Presumably this non-uniqueness is a rather general feature for
arbitrary $s$. However, only one BRST operator for each $s$, for which the
$(\varphi,\beta,\gamma)$ system has the central charge ${2(s-2)\over (s+1)}$
discussed above, seems to be associated with a unitary theory.  It is this
choice that we shall be concentrating on in the present paper.)

\medskip
\noindent{\it 2.2 The spin-2 plus spin-4 string}
\medskip

     Let us consider first the spin-2 plus spin-$4$ string.  The BRST operator
is then given by (2.1)--(2.7), with $\alpha^2=\ft{243}{20}$ and the operator
$F(\varphi,\beta,\gamma)$ given by
$$
\eqalign{
F(\varphi,\beta,\gamma)&=(\del\varphi)^4 + 4\alpha\, \del^2\varphi\,
(\del\varphi)^2 + \ft{41}5 (\del^2\varphi)^2 + \ft{124}{15}
\del^3\varphi\, \del \varphi +\ft{46}{135} \alpha\, \del^4\varphi\cr
& +8 (\del\varphi)^2\, \beta\, \del\gamma -\ft{16}9\alpha\, \del^2\varphi\,
\beta\, \del\gamma -\ft{32}9 \alpha\, \del\varphi\,
\beta\, \del^2 \gamma-\ft45 \beta\, \del^3\gamma + \ft{16}3 \del^2\beta\,
\del\gamma .\cr}\eqno(2.9)
$$

     As usual, physical states $\ket{\chi}$ are determined by the requirement
that they be annihilated by the BRST operator, and that they be BRST
non-trivial.  In other words, $Q_B\ket{\chi}=0$ and $\ket{\chi}\ne
Q_B\ket{\psi}$ for any $\ket{\psi}$.  It was conjectured in [1] and [2] that
all
continuous-momentum physical states for multi-scalar string theories of the
kind
we are discussing can be described by physical operators of the form
$$
V_\Delta=c\, U(\varphi,\beta,\gamma)\, V^{\rm eff}(X),\eqno(2.10)
$$
acting on the $SL(2,C)$ vacuum, where the operator $V^{\rm eff}(X)$ creates an
effective spacetime physical state $\phys_{\rm eff}\equiv V^{\rm
eff}(X(0))\ket{0}$ satisfying the highest-weight conditions
$$
\eqalign{
L^{\rm eff}_n\phys_{\rm eff} &=0,\qquad n>0,\cr
(L^{\rm eff}_0 -\Delta)\phys_{\rm eff}&=0.\cr}\eqno(2.11)
$$
For simplicity, one can always take the effective-spacetime operator
$V^{\rm eff}(X)$ to be tachyonic, since the discussion of physical states with
excitations in the effective spacetime proceeds identically to that of
ordinary string theory.  The interesting new features of the $W$ string
theories are associated with excitations in the $(\varphi,\beta,\gamma)$
fields.  Thus we are primarily concerned with solving for the
operators $U(\varphi,\beta,\gamma)$ that are highest weight under
$T\equiv T_\varphi +T_{\gamma,\beta}$ with conformal weights
$h=1-\Delta$, and that in addition satisfy $[ Q_1,U\}=0$. Solving
these conditions for $U(\varphi,\beta,\gamma)$, with $V^{\rm eff}$
being highest weight under $T^{\rm eff}$ with conformal weight
$\Delta=1-h$, is equivalent to solving the physical-state conditions
for $V_\Delta$ in (2.10).

     In [2], all physical states up to and including level $\ell=9$
in $(\varphi,\beta,\gamma)$ excitations were studied for the spin-2 plus
spin-4 string.  It was found that all the physical states fall into a set
of different sectors, characterised by the value $\Delta$ of the effective
spacetime intercept.  (There is also a sector of discrete physical
states, with zero momentum in the effective spacetime, which will
not be of relevance to us here [2].)  Specifically, for the spin-2 plus spin-4
string, $\Delta$ can take values in the set $\Delta=\{
1,\ft{14}{15},\ft35,\ft13,-\ft25,-2\}$.  As one goes to higher and higher
levels $\ell$, one just encounters repetitions of these same intercept values,
with more and more complicated operators $U(\varphi,\beta,\gamma)$.  These
operators correspondingly have conformal weights $h$ that are conjugate to
$\Delta$, {\it i.e.}\ $h=1-\Delta=\{0, \ft1{15}, \ft25, \ft23, \ft75, 3\}$.
For convenience, we reproduce here the table of results up to level 9, giving
the $(\beta,\gamma)$ ghost number $g$ of the operators
$U(\varphi,\beta,\gamma)$, their conformal weights $h$, and their $\varphi$
momenta $\mu$:

$$
\hbox{
\vbox{\tabskip=0pt \offinterlineskip
\def\tablerule{\noalign{\hrule}}
\halign to350pt{\strut#& \vrule#\tabskip=0em plus10em&
\hfil#\hfil& \vrule#& \hfil#& \vrule#& \hfil#& \vrule#&
\hfil#& \vrule#\tabskip=0pt\cr\tablerule
&&\phantom{}&&${\scriptstyle g}$&&${\scriptstyle h}\ \quad$
&&$\mu\ ({\scriptstyle {\rm In\ units\ of}\ \alpha/27) \quad}$&\cr\tablerule
&&$\ell=0$
&&${\scriptstyle 3}$
&&${\scriptstyle {1\over15}\quad\ 0 }$
&&${\scriptscriptstyle \{-28,-26\}\quad \{-30,-24\}}$&\cr
\tablerule
&&$\ell=1$
&&${\scriptstyle 2}$
&&${\scriptstyle {2\over3}\quad {2\over5} \quad {1\over15}}$
&&${\scriptstyle -20\quad -18\quad -16}$&\cr
\tablerule
&&$\ell=2$
&&${\scriptstyle 2}$
&&${\scriptstyle {7\over5}\ \quad {2\over3}}$
&&${\scriptstyle -18\quad -14}$&\cr
\tablerule
&&$\ell=3$
&&${\scriptstyle 1}$
&&${\scriptstyle {2\over3}\quad {1\over15}}$
&&${\scriptstyle -10\quad -\phantom{1}8}$&\cr
\tablerule
&&$\ell=4$
&&${\scriptstyle 1}$
&&${\scriptstyle {2\over5}}$
&&${\scriptstyle  -\phantom{1}6}$&\cr
\tablerule
&&$\ell=5$
&&${\scriptstyle 1}$
&&${\scriptstyle {7\over5}\ \quad {2\over3}}$
&&${\scriptstyle -\phantom{1}6\quad -\phantom{1}4}$&\cr
\tablerule
&&$\ell=6$
&&${\scriptstyle 0}$
&&${\scriptstyle 0}$
&&${\scriptstyle 0}$&\cr
\tablerule
&&$\ell=7$
&&${\scriptstyle 0}$
&&${\scriptstyle {1\over15}}$
&&${\scriptstyle 2}$&\cr
\tablerule
&&$\ell=8$
&&${\scriptstyle 0}$
&&${\scriptstyle {1\over15}}$
&&${\scriptstyle 4}$&\cr
\tablerule
&&$\ell=9$
&&${\scriptstyle 0}$
&&${\scriptstyle 3\ \quad 0}$
&&${\scriptstyle 0\quad\phantom{-1}6}$&\cr
\tablerule
\noalign{}}}}
$$
\centerline{\it Table 1. $U(\varphi,\beta,\gamma)$ operators for the spin-2
plus spin-4 string}
\bigskip
     The explicit expressions for the operators $U(\varphi,\beta,\gamma)$ can
be
quite complicated, and we shall not give them all here.  Some simple examples
are as follows.  We find $U=c\, \del^2\gamma\, \del\gamma\, \gamma\,
e^{\mu\varphi}$ at level $\ell=0$; $U=c\, \del\gamma\, \gamma\, e^{\mu\varphi}$
at $\ell=1$; $U=\Big( 10\, \del\varphi\, \del\gamma\, \gamma
-(\mu+2\alpha)\del^2\gamma\, \gamma \Big) e^{\mu\varphi}$ at $\ell=2$; and
$U=1$
at $\ell=6$.  The values of the momentum $\mu$ are given in the table.

     We wish to show that the complete set of operators
$U(\varphi,\beta,\gamma)$ can be associated with the highest-weight states of
the lowest $W_3$ minimal model. The key to doing this is to identify the spin-2
and spin-3 currents $T$ and $W$ of the associated $W_3$ algebra, realised on
the
$(\varphi,\beta,\gamma)$ system. For $T$, this is straightforward; it is simply
given by   $$
T=T_\varphi + T_{\gamma,\beta}, \eqno(2.12)
$$
where $T_\varphi$ and $T_{\gamma,\beta}$ are given in (2.4) and (2.5).  For
$W$, we observe from the results in [2] that at level $\ell=9$ there
is an operator $U(\varphi,\beta,\gamma)$ with conformal weight 3, ghost
number $g=0$, and momentum $\mu=0$.  Clearly this is the required primary
spin-3 current.  Its detailed form is
$$
\eqalign{
W=&\sqrt{\ft2{13}} \Big\{\ft53\, (\del\varphi)^3 +5\alpha
\,\del^2\varphi\, \del\varphi +\ft{25}4\, \del^3\varphi + 20\, \del\varphi\,
\beta\, \del\gamma\cr
&\qquad+12 \, \del\varphi\, \del\beta\, \gamma + 12\, \del^2\varphi\,
\beta\, \gamma + 5\alpha \, \del\beta\, \del \gamma + 3\alpha
\, \del^2\beta\, \gamma\Big\}, \cr}\eqno(2.13)
$$
where we have given it the canonical normalisation in which $W(z) W(w)\sim
{c/3\over (z-w)^6}+ \hbox{more}$, with the central charge $c=\ft45$.  It is
now a straightforward matter to compute the OPEs of the $T$ and $W$ currents
and verify that they do indeed generate the $W_3$ algebra at $c=\ft45$.  The
only noteworthy point in the verification is that at the second-order pole in
the OPE of $W$ with $W$ there is an additional spin-4 primary current $H$
over and above the expected terms.  However it turns out that $H$ is
null, and so it does not upset the fact that $T$ and $W$ generate the
$W_3$ algebra.  In fact, $H$ is the BRST-trivial current $\{Q_1,\beta
\}$.

     Having found the currents that generate the $W_3$ algebra, we are
now in a position to see how they act on the operators
$U(\varphi,\beta,\gamma)$ of the physical states of the spin-2 plus
spin-4 string.  Of course we already know that the operators
$U(\varphi,\beta,\gamma)$ are primary fields under $T$.  Acting with $W$, we
find that when $h$ takes values in the set $\{0,\ft1{15}, \ft25, \ft23\}$, the
corresponding operators are highest-weight under $W$, {\it i.e.}\
$[W_n,U(0)]=0$ for $n >0$, and $[W_0,U(0)]=\omega\, U(0)$.  We find that the
weights of the highest-weight operators are as follows\footnote{$^*$}
{\tenfoot To be precise, we find that for
$h=\ft1{15}$ the $W_0$ weight is positive for those operators
$U(\varphi,\beta,\gamma)$ that have
$({\alpha\over 27})^{-1}\mu= 4\,
\hbox{mod}\, 6$, and negative when $({\alpha\over 27})^{-1}\mu=
2\,\hbox{mod}\, 6$.  Similarly, for operators with $h=\ft23$, we find the
$W_0$ weight is positive when $({\alpha\over 27})^{-1}\mu= 2\,
\hbox{mod}\, 6$, and negative when $({\alpha\over 27})^{-1}\mu=
4\, \hbox{mod}\, 6$.  These results accord with the observation in [2] that
there are two independent towers of $h=\ft1{15}$ operators, and two independent
towers of $h=\ft23$ operators, with the screening operator $\beta\, e^{\ft29
\alpha\varphi}$ generating each tower from its lowest-level member.}:
$$
\eqalign{
L_0:&\qquad \{0,\ft1{15},\ft25,\ft23\},\cr
\cr
{243\over\alpha}\sqrt{\ft{13}8} \, W_0 :&\qquad \{0,\pm1,0,\pm26\}.\cr}
\eqno(2.14)
$$
Comparing with the results in [7], we see that these $L_0$ and $W_0$ weights
are
precisely those for the lowest $W_3$ minimal model, with $c=\ft45$.  The
remaining operators $U(\varphi,\beta,\gamma)$ in the physical states of the
spin-2 plus spin-4 string have $L_0$ weights $h=\ft75$ and 3.  We find that
these are not highest weight under the $W$ current.  In fact, they are $W$
descendant fields; those with $h=\ft75$ can be written as $W_{-1}+\cdots$
acting
on operators $U(\varphi,\beta,\gamma)$ with $h=\ft25$, and those with $h=3$ can
be written as $W_{-3}+\cdots$ acting on operators $U(\varphi,\beta,\gamma)$
with
$h=0$.

     The conclusion of the above discussion is that the
$U(\varphi,\beta,\gamma)$ operators appearing in the physical states of the
spin-2 plus spin-4 string are precisely those associated with the $c=\ft45$
lowest $W_3$ minimal model.  Those with $h=\{0,\ft1{15}, \ft25, \ft23 \}$ are
$W_3$ highest-weight fields, whilst those with $h=\ft75$ and 3 are $W_3$
descendants.  Viewed as purely Virasoro fields, they are all primaries.  In
fact, what we are seeing is an explicit example of the phenomenon under which
the set of highest-weight fields of a $W$ minimal model decomposes into a
larger set of highest-weight fields with respect to the Virasoro subalgebra.
In this example, since $c$ is less than 1, the $W$ fields decompose into a {\it
finite} number of Virasoro primaries (namely a subset of the primaries of the
$c=\ft45$ 3-state Potts model).  In a more generic example, where the $W$
minimal model has $c\ge 1$, the finite number of $W$ highest-weight states will
decompose into an infinite number of Virasoro primaries, with infinitely many
of them arising as $W$ descendants.  We shall encounter explicit examples of
this when we study the spin-2 plus spin-$s$ strings with $s=5$ and $s=6$.

\np
\noindent{\it 2.3 The spin-2 plus spin-5 string}
\medskip

     Let us now turn to the example of the spin-2 plus spin-5 string.
The operator $F(\varphi,\beta,\gamma)$ appearing in (2.3) is now given by [2]
$$
\eqalign{
F(\beta,\gamma,\varphi)&=(\del\varphi)^5 +5\alpha\, \del^2\varphi\,
(\del\varphi)^3+\ft{305}8 (\del^2\varphi)^2\, \del\varphi
+\ft{115}6 \del^3\varphi\, (\del\varphi)^2 +\ft{10}3 \alpha\,
\del^3\varphi\, \del^2\varphi\cr
&+\ft{55}{48}\alpha\, \del^4\varphi\, \del\varphi
+\ft{251}{576} \del^5\varphi +\ft{25}2 (\del\varphi)^3\,\beta\,
\del\gamma +\ft{25}4 \alpha\, \del^2\varphi\, \del\varphi \, \beta\,
\del\gamma + \ft{25}4 \alpha\, (\del\varphi)^2\, \del\beta\,\del\gamma\cr
&+\ft{125}{16} \del^3\varphi\, \beta\, \del\gamma +\ft{325}{12}
\del^2\varphi\, \del\beta\, \del\gamma +\ft{375}{16} \del\varphi\,
\del^2\beta\, \del\gamma-\ft{175}{48}\del\varphi\, \beta\, \del^3\gamma
\cr
&+\ft53 \alpha\, \del^3\beta\, \del\gamma -\ft{35}{48}\alpha\, \del\beta\,
\del^3\gamma,\cr}\eqno(2.15)
$$
with $\alpha^2=\ft{121}6$.
Here, we have $c^{\rm eff}=25$, and the associated minimal model, with
$c=1$, is expected to be the lowest $W_4$ minimal model [2].  Following
the same strategy as before, we should therefore begin by looking
amongst the operators $U(\varphi,\beta,\gamma)$ associated with the
physical states for examples with zero $\varphi$ momentum, and ghost
number $g=0$, at spins 3 and 4.  These will be the candidate spin 3
and spin 4 primary fields of the $W_4$ algebra.  In [2], all physical
states of the spin-2 plus spin-5 string up to level $\ell=13$ were
obtained.  In fact one can easily see that the required physical
states associated with the spin-3 and spin-4 currents will occur at
levels 13 and 14 respectively.  Thus we seek such physical states of the
form (2.10), with $U(\varphi,\beta,\gamma)$ having zero ghost number and
zero $\varphi$ momentum.  In other words, $U(\varphi,\beta,\gamma)$
should be primary under $T=T_\varphi + T_{\gamma,\beta}$, with weight 3
or 4 respectively, and satisfy $\{Q_1, U\}=0$.  We find the
following expressions for the primary spin-3 current $W$ and spin-4
current $V$ of the $W_4$ algebra at $c=1$:
$$
\eqalignno{
W&=\ft12 (\del\varphi)^3 + \ft32\alpha\,
\del^2\varphi\, \del\varphi +\ft{31}{12}\, \del^3\varphi +
\ft{15}2\, \del\varphi\, \beta\, \del\gamma\cr
&+5\, \del\varphi\, \del\beta\, \gamma + 5\, \del^2\varphi\,
\beta\, \gamma + \ft32\alpha\, \del\beta\, \del \gamma +
\alpha\, \del^2\beta\, \gamma , &(2.16)\cr
\cr
V&=-{25\over \sqrt{864}}\Big\{ (\del\varphi)^4 + 4\alpha\,
\del^2\varphi\, (\del\varphi)^2 + \ft{2317}{150}\, (\del^2\varphi)^2 +
\ft{277}{25}\, \del^3\varphi\, \del\varphi + \ft{617}{1650}\alpha\,
\del^4\varphi\cr
&\qquad+\ft{108}{5}\,\del^2\varphi\, \del\varphi\, \beta\, \gamma
 +20\, (\del\varphi)^2\, \beta\, \del\gamma +\ft{292}{25}\,
(\del\varphi)^2\, \del\beta\, \gamma +\ft{62}{11}\alpha\, \del^2\varphi\,
\beta\, \del\gamma \cr
&\qquad+ \ft{2104}{275}\alpha\, \del^2\varphi\,
\del\beta\, \gamma
+ \ft{216}{55}\alpha\, \del\varphi\, \del^2\beta\, \gamma +\ft{378}{55}
\alpha \,\del\varphi\, \del\beta\, \del\gamma + \ft{108}{55} \alpha\,
\del^3\varphi\, \beta\, \gamma \cr
&\qquad-\ft{44}{15}\, \beta\, \del^3\gamma
-\ft{132}{25}\, \del\beta\, \del^2\gamma +\ft{321}{25}\, \del^2\beta\,
\del\gamma + \ft{544}{75}\, \del^3\beta\, \gamma
+\ft{44}5\, \del\beta\, \beta\, \del\gamma\, \gamma\Big\}.
&(2.17)\cr}
$$
We have normalised these currents canonically, so that the coefficient
of the highest-order pole in the OPE of a spin-$s$ current with itself is
$c/s$, where the central charge is $c=1$ in the present case.

     It is now a straightforward matter to check that $T$, $W$ and $V$
indeed generate the $W_4$ algebra at $c=1$.  We find complete
agreement with the algebra given in [8], again modulo the appearance of
certain additional primary fields that are BRST exact, and hence null.
Specifically, we find a spin-5  null primary field $\{Q_1,\beta\}$
and its Virasoro descendants in the OPE of $W$ with $V$, and
a spin-6 null primary field $\{Q_1,(30\,\del\varphi\,\beta + 11 \sqrt{6}\,
\del\beta)\}$ and its descendants in the OPE of $V$ with $V$.

      Having obtained the currents that generate the $W_4$ algebra, we
may now examine the $U(\varphi,\beta,\gamma)$ operators in the
physical states of the spin-2 plus spin-5 string, in order to compare
their weights with those of the lowest $W_4$ minimal model.
Specifically, this model has primary fields with conformal weights
$h=\{ 0, \ft1{16}, \ft1{12}, \ft13,\ft9{16},\ft34,1 \}$.  The results
presented in [2], extended to level $\ell=14$, are given in Table 2
below.
$$
\hbox{
\vbox{\tabskip=0pt \offinterlineskip
\def\tablerule{\noalign{\hrule}}
\halign to400pt{\strut#& \vrule#\tabskip=0em plus20em&
\hfil#\hfil& \vrule#& \hfil#& \vrule#& \hfil#& \vrule#&
\hfil#& \vrule#\tabskip=0pt\cr\tablerule
&&\phantom{}&&${\scriptstyle g}$&&${\scriptstyle h}\qquad\qquad$
&&$\mu\ {\scriptstyle ({\rm In\ units\ of}\ \alpha/22)}$\qquad&\cr
\tablerule
&&$\ell=0$
&&${\scriptstyle 4}$
&&${\scriptstyle {1\over 12}}\quad\phantom{-}{\scriptstyle {1\over 16}}
\quad{\scriptstyle \phantom{1}0}$
&&${\scriptstyle -22}\quad
{\scriptscriptstyle \{-23, -21\}}\quad {\scriptscriptstyle \{-24,-20\}}$&\cr
\tablerule
&&$\ell=1$
&&${\scriptstyle 3}$
&&${\scriptstyle {3\over 4}}\quad{\scriptstyle \phantom{-}{9\over 16}
\quad {\scriptstyle \phantom{-1}{1\over 3}}
\quad {\scriptstyle {1\over16}}}$
&&${\scriptstyle -18}\quad {\scriptstyle -17}\quad {\scriptstyle -16}\quad
{\scriptstyle -15}$&\cr
\tablerule
&&$\ell=2$
&&${\scriptstyle 3}$
&&${\scriptstyle {25\over 16}}\quad \ {\scriptstyle
{4\over 3}}\quad{\scriptstyle \phantom{1}{3\over 4}}$
&&${\scriptstyle -17}\quad
{\scriptstyle -16}\quad {\scriptstyle -14}$&\cr
\tablerule
&&$\ell=3$
&&${\scriptstyle 2}$
&&${\scriptstyle {\scriptstyle \phantom{-}1}\, \quad\phantom{-}{9\over16}}
\quad{\scriptstyle {1\over 12}}$
&&${\scriptstyle -12}\quad {\scriptstyle -11}\quad {\scriptstyle -10}
$&\cr
\tablerule
&&$\ell=4$
&&${\scriptstyle 2}$
&&${\scriptstyle {25\over16}}\quad {\scriptstyle {9\over16}}$
&&${\scriptstyle -11}\quad {\scriptstyle  -\phantom{1}9} $&\cr
\tablerule
&&$\ell=5$
&&${\scriptstyle 2}$
&&${\scriptstyle 3 }\quad{\scriptstyle {25\over12} }\quad
{\scriptstyle {25\over16}} \quad {\scriptstyle \phantom{1}1}$
&&${\scriptstyle -12} \quad {\scriptstyle -10} \quad
 {\scriptstyle -\phantom{1}9 \quad {\scriptstyle -\phantom{1}8}} $&\cr
\tablerule
&&$\ell=6$
&&${\scriptstyle 1}$
&&${\scriptstyle {3\over4}} \quad{\scriptstyle {1\over16}}$
&&${\scriptstyle -\phantom{1}6} \quad {\scriptstyle -\phantom{1}5} $&\cr
\tablerule
&&$\ell=7$
&&${\scriptstyle 1}$
&&${\scriptstyle \phantom{-}{1\over3}}$
&&${\scriptstyle -\phantom{1}4} $&\cr
\tablerule
&&$\ell=8$
&&${\scriptstyle 1}$
&&${\scriptstyle {4\over3}} \quad{\scriptstyle {9\over16}}$
&&${\scriptstyle -\phantom{1}4} \quad {\scriptstyle -\phantom{1}3} $&\cr
\tablerule
&&$\ell=9$
&&${\scriptstyle 1}$
&&${\scriptstyle {49\over16}} \quad{\scriptstyle {25\over16}}
\quad {\scriptstyle \phantom{1}{3\over4}}$
&&${\scriptstyle -\phantom{1}5}\quad {\scriptstyle -\phantom{1}3} \quad
{\scriptstyle -\phantom{1}2} $&\cr
\tablerule
&&$\ell=10$
&&${\scriptstyle 0}$
&&${\scriptstyle 0}$
&&${\scriptstyle 0}
$&\cr
\tablerule
&&$\ell=11$
&&${\scriptstyle 0}$
&&${\scriptstyle {1\over16}}$
&&${\scriptstyle 1}$&\cr
\tablerule
&&$\ell=12$
&&${\scriptstyle 0}$
&&${\scriptstyle {1\over12}}$
&&${\scriptstyle 2}$&\cr
\tablerule
&&$\ell=13$
&&${\scriptstyle 0}$
&&${\scriptstyle 3} \quad{\scriptstyle {1\over16}}$
&&${\scriptstyle 0}\quad {\scriptstyle \phantom{-1}3}$&\cr
\tablerule
&&$\ell=14$
&&${\scriptstyle 0}$
&&${\scriptstyle 4}\quad{\scriptstyle \phantom{-}{49\over 16}
\quad {\scriptstyle \phantom{}{25\over 12}}
\quad {\scriptstyle \phantom{1}0}}$
&&${\scriptstyle 0}\quad {\scriptstyle \phantom{-1}1}\quad {\scriptstyle
\phantom{-1}2}\quad
{\scriptstyle\phantom{-1} 4}$&\cr
\tablerule
\noalign{}}}}
$$
\centerline{\it Table 2. $U(\varphi,\beta,\gamma)$ operators for the spin-2
plus spin-5 string}
\bigskip
    One can see from the results in Table 2 that indeed all
the conformal weights of the primary fields of the lowest $W_4$
minimal model occur in the spin-2 plus spin-5 string.  We find that the
corresponding weights under the $W_4$ currents (2.16) and (2.17) are:
$$
\eqalign{
L_0:&\qquad \{0,\ft1{16},\ft1{12},\ft13,\ft9{16},\ft34,1\},\cr
\cr
{352\over\alpha}\, W_0:&\qquad \{0,\pm 1, 0,0, \pm11, \pm32, 0\},\cr
\cr
6912\sqrt{6}\, V_0:&\qquad \{0,27,-64,128,-405,1728,-6912\}.\cr}
\eqno(2.18)
$$
We have checked that these weights agree with those that one finds using the
highest-weight vertex-operators of the $W_N$ minimal models in the ``Miura''
realisations discussed in [9], after converting from the non-primary basis of
Miura currents to the primary basis that we are using
here.\footnote{$^*$}{\tenfoot If $h=\ft1{16}$, the weight under the spin-3
current $W$ is positive when $({\alpha\over 22})^{-1}\mu= 3\ \hbox{mod}\ 4$,
and negative when $({\alpha\over 22})^{-1}\mu= 1\ \hbox{mod}\ 4$.  If
$h=\ft9{16}$, the weight under the spin-3 current $W$ is positive when
$({\alpha\over 22})^{-1}\mu= 1\ \hbox{mod}\ 4$, and negative when
$({\alpha\over 22})^{-1}\mu= 3\ \hbox{mod}\ 4$.  If $h=\ft34$, for which
$({\alpha\over22})^{-1}\mu =12 n -18$ or $12n-14$, with $n$ a non-negative
integer (see [2]), the spin-3 weight is positive when $n$ is odd, and negative
when $n$ is even.}  The remaining $U(\varphi,\beta,\gamma)$ operators
obtained here and in [2], with conformal weights $h$ that lie outside the set
of weights for the $W_4$ minimal model, correspond to $W$ and $V$ descendant
states.  In other words, they are secondaries of the $W_4$ minimal model, but
they are primaries with respect to a purely Virasoro $c=1$ model.  In this
more generic case, with $c\ge 1$, the number of primaries in the purely
Virasoro model will be infinite.  Thus if we would go on solving the
physical-state conditions at higher and higher levels $\ell$, we would find a
set of operators $U(\varphi,\beta,\gamma)$ with conformal weights $h$ that
increased indefinitely.  All those lying outside the set $h=\{0,
\ft1{16},\ft1{12},\ft13,\ft{9}{16},\ft34,1\}$ would be given by certain
integers added to values lying in the set, corresponding to $W$ and $V$
descendant fields.

\medskip
\noindent{\it 2.4 The spin-2 plus spin-6 string}
\medskip

     In [2], it was found that there are four different nilpotent BRST
operators of the form (2.1)--(2.7), corresponding to different values
of $\alpha$, and hence $c^{\rm eff}$.  As usual, we shall be concerned
with the case which seems to be associated with a unitary string
theory.  This is given by $\alpha^2=\ft{845}{20}$, implying
$c^{\rm eff}=\ft{174}7$ and hence the $(\varphi,\beta,\gamma)$ system
describes a model with $c=\ft87$.  We expect this to be the lowest
$W_5$ minimal model.
The operator $F(\varphi,\beta,\gamma)$ in this
case takes the form [2]:
$$
\eqalign{
F(\beta,\gamma,\varphi)&=(\del\varphi)^6+ 6\alpha\, \del^2\varphi\,
(\del\varphi)^4 +\ft{765}7 (\del^2\varphi)^2\, (\del\varphi)^2+
\ft{256}7 \del^3\varphi\, (\del\varphi)^3 +\ft{174}{35}\alpha\,
(\del^2\varphi)^3\cr
&+\ft{528}{35}\alpha\, \del^3\varphi\, \del^2\varphi\, \del\varphi +\ft{18}7
\alpha\, \del^4\varphi\, (\del\varphi)^2 +\ft{1514}{245} (\del^3\varphi)^2
+\ft{2061}{245} \del^4\varphi\, \del^2\varphi +\ft{2736}{1225}
\del^5\varphi\, \del\varphi\cr
&+\ft{142}{6125}\alpha\, \del^6\varphi +18(\del\varphi)^4\, \beta\del\gamma +
\ft{72}{5}\alpha\, \del^2\varphi\, (\del\varphi)^2\, \beta\, \del\gamma
+\ft{48}5 \alpha\, (\del\varphi)^3\, \del\beta\, \del\gamma \cr
&+\ft{216}5
\del^3\varphi\, \del\varphi\, \beta\,\del\gamma
+\ft{1494}{35} (\del^2
\varphi)^2 \,\beta\,\del\gamma +\ft{5256}{35} \del^2\varphi\,
\del\varphi\, \del\beta\, \del\gamma+\ft{324}5 (\del\varphi)^2
\,\del^2\beta\,\del\gamma\cr
&-\ft{72}7 (\del\varphi)^2\, \beta\,\del^3\gamma
+\ft{204}{175}\alpha\, \del^4\varphi\,\beta\, \del\gamma +\ft{192}{25}\alpha\,
\del^3\varphi\, \del\beta\, \del\gamma+\ft{2376}{175}\alpha\, \del^2\varphi\,
\del^2\beta\, \del\gamma\cr
&-\ft{144}{175}\alpha\, \del^2\varphi\, \beta\, \del^3\gamma
+\ft{1296}{175}\alpha\, \del\varphi\, \del^3\beta\,
\del\gamma-\ft{576}{175}\alpha\, \del\varphi\, \del\beta\,
\del^3\gamma+\ft{1614}{175} \del^4\beta\, \del\gamma\cr
&-\ft{216}{35}
\del^2\beta\, \del^3\gamma +\ft{144}{1225}
\beta\,\del^5\gamma+\ft{144}{35} \del\beta\,\beta\,
\del^2\gamma\,\del\gamma.\cr}\eqno(2.19)
$$

     In [2], physical states in this theory up to and including level
$\ell=6$ were studied.  Here, we are primarily concerned with finding
the physical states associated with the expected spin-3, spin-4 and
spin-5 primary fields of the $W_5$ minimal model.  These should occur
at levels $\ell=18$, 19 and 20 respectively.  It is a straightforward
matter to solve for such physical states corresponding to operators
$U(\varphi,\beta,\gamma)$ with zero ghost number, and zero $\varphi$
momentum.
We find the following results for the spin-3, spin-4 and
spin-5 operators $W$, $V$ and $Y$:
$$
\eqalign{
W&=\sqrt{\ft2{57}}\Big\{\ft73 (\del\varphi)^3 + 7\alpha\,
\del^2\varphi\, \del\varphi +\ft{185}{12}\, \del^3\varphi +
42\, \del\varphi\, \beta\, \del\gamma\cr
&\qquad +30 \, \del\varphi\, \del\beta\, \gamma + 30 \, \del^2\varphi\,
\beta\, \gamma + 7\alpha\, \del\beta\, \del \gamma +
 5\alpha\, \del^2\beta\, \gamma\Big\} ,\cr} \eqno(2.20)
$$
\np
$$
\eqalignno{
V&=-\sqrt{\ft7{60819}}\Big\{\ft{427}8\, (\del\varphi)^4 + \ft{427}2\alpha\,
\del^2\varphi\, (\del\varphi)^2 + \ft{10619}{8}\, (\del^2\varphi)^2 +
743\, \del^3\varphi\, \del\varphi + \ft{3313}{156}\alpha\,
\del^4\varphi\cr
&\qquad+1455\,\del^2\varphi\, \del\varphi\, \beta\, \gamma
 +1281\, (\del\varphi)^2\, \beta\, \del\gamma +825 \,
(\del\varphi)^2\, \del\beta\, \gamma +\ft{5370}{13}\alpha\, \del^2\varphi\,
\beta\, \del\gamma \cr
&\qquad+ \ft{6900}{13}\alpha\, \del^2\varphi\,
\del\beta\, \gamma
+ \ft{2910}{13}\alpha\, \del\varphi\, \del^2\beta\, \gamma +\ft{4656}{13}
\alpha \,\del\varphi\, \del\beta\, \del\gamma + \ft{1455}{13} \alpha\,
\del^3\varphi\, \beta\, \gamma \cr
&\qquad-247\, \beta\, \del^3\gamma
-494\, \del\beta\, \del^2\gamma +\ft{6891}{7}\, \del^2\beta\,
\del\gamma +\ft{7785}{14}\, \del^3\beta\, \gamma
+1170\, \del\beta\, \beta\, \del\gamma\, \gamma\Big\}.
&(2.21)\cr
\cr
Y&=\sqrt{\ft{7}{122}}\Big\{ \ft{749}{165}\, (\del\varphi)^5 +
\ft{749}{33}\alpha\, (\del\varphi)^3\, \del^2\varphi + \ft{6091}{22}\,
(\del^2\varphi)^2\, \del\varphi + \ft{1361}{66}\alpha\,
\del^3\varphi\, \del^2\varphi\cr
&\qquad +\ft{14351}{132}\, \del^3\varphi\, (\del\varphi)^2\, +
\ft{2330}{429}\alpha\, \del^4\varphi\, \del\varphi  +\ft{4825}{1848}\,
\del^5\varphi + \ft{1498}{11}(\del\varphi)^3\, \beta\, \del\gamma \cr
&\qquad +\ft{2570}{33}\, (\del\varphi)^3\, \del\beta\, \gamma  +
\ft{11382}{143}\alpha\, \del^2\varphi\, \del\varphi\, \beta\,
\del\gamma + \ft{15670}{143}\alpha\, \del^2\varphi\, \del\varphi\,
\del\beta\, \gamma + \ft{430}{11}\alpha\, (\del^2\varphi)^2\, \beta\,
\gamma \cr
&\qquad +\ft{4270}{143}\alpha\, \del^3\varphi\, \del\varphi\, \eta\,
\gamma +\ft{2350}{11}\, \del^2\varphi\, (\del\varphi)^2\, \beta\,
\gamma + \ft{7840}{143}\alpha\, (\del\varphi)^2\, \del\beta\,
\del\gamma +\ft{4380}{143}\alpha\, (\del\varphi)^2\, \del^2\beta\,
\gamma \cr
&\qquad -52\, \del\varphi\, \beta\, \del^3\gamma -\ft{624}{7}\,
\del\varphi\, \del\beta\, \del^2\gamma + \ft{17331}{77}\,
\del\varphi\, \del^2\beta\, \del\gamma +\ft{9775}{77}\, \del\varphi\,
\del^3\beta\, \gamma \cr
&\qquad -\ft{624}{7}\, \del^2\varphi\, \beta\, \del^2\gamma +
\ft{18541}{77} \, \del^2\varphi\, \del\beta\, \del\gamma +
\ft{3390}{11}\, \del^2\varphi\, \del^2\beta\, \gamma+ \ft{7957}{154}\,
\del^3\varphi\, \beta\, \del\gamma\cr
&\qquad +\ft{77705}{462}\, \del^3\varphi\,\del\beta\, \gamma
+\ft{11575}{462}\, \del^4\varphi\, \beta\, \gamma -\ft{26}{3}\alpha\,
\del\beta \, \del^3\gamma -\ft{104}{7}\alpha\, \del^2\beta\,
\del^2\gamma \cr
&\qquad +\ft{6775}{1001}\alpha\, \del^3\beta\, \del\gamma
+\ft{5365}{1001} \alpha\, \del^4\beta\, \gamma +120\, \del\varphi \,
\del\beta\, \beta\, \del\gamma\, \gamma +\ft{120}{13}\alpha\,
\del^2\beta\, \beta\, \del\gamma\, \gamma\Big\}. &(2.22)\cr}
$$
We have as usual given these currents their canonical normalisations.
We have checked that they indeed, together with $T=T_\varphi +
T_{\gamma,\beta}$, generate the $W_5$ algebra, given in [10], with
central charge $c=\ft87$.
Again, one finds additional BRST exact fields appearing on the right-hand
sides of the OPEs of the primary currents.  These fields are primaries (and
their descendants) except in the case of the OPE $Y(z) Y(w)$. The new field
occuring at the second order pole of this OPE is only primary up to BRST
exact terms.

     It was found in [2] that the physical states of the spin-2 plus
spin-6 string were associated with operators $U(\varphi,\beta,\gamma)$
whose conformal weights included those of the highest-weight fields of the
lowest $W_5$ minimal model, which has $c=\ft87$. Indeed, here we find that the
highest-weight fields have the weights
$$
\eqalign{
L_0:&\qquad \{0,\ft2{35},\ft3{35}, \ft27, \ft{17}{35},\ft{23}{35},
\ft45, \ft67,\ft65 \},\cr
\cr
{325\over\alpha} \sqrt{\ft{57}{8}}\, W_0:&\qquad \{0,\pm2, \pm1, 0,
\pm13, \pm39, \pm76, 0, \pm38\} ,\cr
\cr
25\sqrt{\ft{141911}3}\, V_0:&\qquad \{0,11,-14, 50, -74, 66, 836, -1100,
-2299\},\cr
\cr
{89375\over\alpha} \sqrt{\ft{427}{32} }\, Y_0:&\qquad \{ 0,\pm 11, \mp
48, 0, \pm314, \mp 902, \pm 1452, 0, \mp16621\}.\cr}\eqno(2.23)
$$
Note that the $\pm$ signs for the weights under the spin-3 current
$ W$ are correlated with those for the weights under the
spin-5 current $Y$.  Again we have checked that these weights agree with those
calculated from the realisations of the $W_N$ minimal models given in [9].
All the physical states of the
spin-2 plus spin-6 string are presumably associated with operators
$U(\varphi, \beta, \gamma)$ that are either highest-weight under the
$W_5$ algebra, as given in (2.23), or they are $W$, $V$ or $Y$
descendants of such operators.  Some examples of descendant operators
were found in [2].  Again one expects, since the $W_5$ minimal model
has $c=\ft87 \ge1$, that there will be an infinite number of descendant
operators.

\bigskip
\noindent{\bf 3. Discussion}
\bigskip

     In this paper, we have investigated the physical states of a
string theory based on constraints imposed by holomorphic currents of
spins 2 and $s$.  In [2], where these models were first proposed, it
was conjectured that the physical states of the spin-2 plus spin-$s$
string are described by effective Virasoro strings whose spacetime
energy-momentum tensor has central charge $26-\ft{2(s-2)}{(s+1)}$.
The physical states factorise as in (2.10), with the intercepts
$\Delta$ for the effective-spacetime states being conjugate to the
conformal weights $h$ of the operators $U(\varphi,\beta,\gamma)$, in
the sense that $\Delta=1-h$.  It was conjectured in [2] that the
$(\varphi,\beta,\gamma)$ system, which has central charge
$c=\ft{2(s-2)}{(s+1)}$, should describe the lowest unitary
minimal model of the $W_{s-1}$ algebra.

     We have tested the above conjecture in detail for the cases $s=4$, 5 and 6
of the spin-2 plus spin-$s$ string.  We have shown, for the lowest few levels,
that indeed the operators $U(\varphi,\beta,\gamma)$ that arise in the physical
states (2.10) are associated with the highest-weight fields of the lowest
unitary $W_{s-1}$ minimal models.  Specifically, we find in all physical states
(2.10) that $U(\varphi,\beta,\gamma)$ is either a highest-weight field of the
corresponding $W_{s-1}$ minimal model, or else it is a descendant field in the
sense that it is obtained from a highest-weight field by acting with the
negative modes of the primary currents of the $W_{s-1}$ algebra.  Since the
central charge $c=\ft{2(s-2)}{(s+1)}$ of the $(\varphi,\beta,\gamma)$ system
satisfies $c\ge1$ for $s\ge5$, it follows that in these cases there are
infinite numbers of such descendant fields in the models.  Thus the $W_{s-1}$
currents provide a strikingly powerful organising symmetry in these cases.

     The original realisations of the $W_N$ algebras were the
$(N-1)$-scalar realisations from the Miura transformation, introduced
in [7,9].  By contrast, the realisations of the lowest unitary
$W_{s-1}$ minimal models that we find here are all given in terms of
just one scalar field $\varphi$, and the $(\beta,\gamma)$ ghost system
for spin-$s$.  This ghost system can be bosonised, yielding two-scalar
realisations. However even when $s=4$, our two-scalar realisation is
quite different from the usual Miura realisation of $W_3$.  In
particular, our realisations close on the $W_{s-1}$ algebras modulo
the appearance of certain null primary fields in the OPEs of the
currents, whereas no such null fields arise in the Miura realisations.
Presumably the realisations that we find here are very
specific to the particular unitary minimal models that arise in these
higher-spin string theories.  As an example, we present the spin-2 and
spin-3 currents (2.12) and (2.13) for the $W_3$ algebra at $c=\ft45$
in the bosonised language, where $\gamma= :e^{i\rho}:$ and
$\beta=:e^{-i\rho}:$.
$$
\eqalign{
T&=-\ft12 (\del\varphi)^2 -\ft12 (\del\rho)^2 -\alpha\, \del^2\varphi
+\ft72 i\, \del^2\rho,\cr
\cr
W&=\sqrt{\ft2{13}}\Big\{ \ft53\,(\del\varphi)^3 +5\alpha\,
\del^2\varphi\, \del\varphi +\ft{25}4\, \del^3\varphi +4\,
\del\varphi\, (\del\rho)^2 -16i\, \del\varphi\, \del^2\rho \cr
&\qquad\qquad -12i\, \del^2\varphi\, \del\rho -\ft23 i\alpha\, (\del\rho)^3
-3\alpha\, \del^2\rho\, \del\rho -\ft{11}6 i\alpha\, \del^3\rho\Big\},\cr}
\eqno(3.1)
$$
where $\alpha^2=\ft{243}{20}$.  It is interesting to note that this realisation
of the $W_3$ algebra at $c=\ft45$ is precisely the one obtained in [11] (case
$I$, after an $SO(1,1)$ rotation of the two scalars), where more general scalar
realisations of $W_3$ modulo a null spin-4 operator were considered.

     Finally, we remark that, as observed in [2], the spectrum of
physical states for the spin-2 plus spin-$s$ string becomes more
complicated if there is just one $X^\mu$ coordinate in the effective
energy-momentum tensor $T^{\rm eff}$ (2.7).  In particular, there are
additional physical states over and above those of the form (2.10),
which do not factorise into the product of effective-spacetime
physical states times operators $U(\varphi,\beta,\gamma)$.  Examples
of these were found for the $W_3$ string in [12], and for spin-2 plus
spin-$s$ strings in [2].  A general discussion of the BRST cohomology
for the two-scalar $W_3$ string is given in [13].  It may well be that
the spin-2 plus spin-$s$ strings with just one additional coordinate
$X^\mu$ capture the more subtle aspects of the underlying higher-spin
geometry.

\bigskip\bigskip
\centerline{\bf ACKNOWLEDGMENTS}
\bigskip

     We are grateful to Horst Kausch for discussions, and to Klaus
Hornfeck for supplying us with files containing his results for the
$W_4$ and $W_5$ algebras.  K.T. is grateful to Texas A\&M for  hospitality,
and C.N.P. thanks Imperial College, London for hospitality during the course
of this work.

\np
\singlespace
\centerline{\bf REFERENCES}
\frenchspacing
\bigskip

\item{[1]}H. Lu, C.N. Pope, S. Schrans and X.J. Wang,  ``On the spectrum
and scattering of $W_3$ strings,'' preprint CTP TAMU-4/93, KUL-TF-93/2,
hep-th/9301099, to appear in {\sl Nucl. Phys.} {\bf B}.

\item{[2]}H. Lu, C.N. Pope and X.J. Wang,  ``On higher-spin
generalisations of string theory,'' preprint CTP TAMU-22/93,
hep-th/9304115.

\item{[3]}M. Bershadsky, W. Lerche, D. Nemeschansky and N.P. Warner,
{\sl Phys. Lett.} {\bf B292} (1992) 35.

\item{[4]}E. Bergshoeff, A. Sevrin and X. Shen, {\sl Phys. Lett.} {\bf
B296} (1992) 95.

\item{[5]}M. Bershadsky, W. Lerche, D. Nemeschansky and N.P. Warner,
``Extended $N=2$ superconformal structure of gravity and $W$-gravity
coupled to matter,'' USC-92/021, CERN-TH.6694/92.

\item{[6]}K. Thielemans, {\sl Int. J. Mod. Phys.} {\bf C2} (1991) 787.

\item{[7]}V.A. Fateev and A.B. Zamolodchikov, {\sl Nucl. Phys.} {\bf B280}
(1987) 644.

\item{[8]}R. Blumenhagen, M. Flohr, A. Kliem, W. Nahm, A. Recknagel
and R. Varnhagen, {\sl Nucl. Phys.} {\bf B361} (1991) 255.\nl
H.G. Kausch and G.M.T. Watts, {\sl Nucl. Phys.} {\bf B354} (1991) 740.

\item{[9]}V.A. Fateev and S.L. Lukyanov, {\sl Int. J. Mod. Phys.} {\bf
A3} (1988) 507.

\item{[10]}K. Hornfeck, ``$W$ algebras with a set of primary fields of
dimensions (3,4,5) and (3,4,5,6),'' KCL-TH-92-9, hep-th/9212104.

\item{[11]}E. Bergshoeff, H.J. Boonstra and M. de Roo, ``Realisations of $W_3$
symmetry,'' preprint, UG-6/92, hep-th9209065.

\item{[12]}C.N. Pope, E. Sezgin, K.S. Stelle and X.J. Wang, {\sl Phys. Lett.}
{\bf B299} (1993) 247.

\item{[13]}P. Bouwknegt, J. McCarthy and K. Pilch, ``Semi-infinite
cohomology of $W$ algebras,'' USC-93/11, hep-th9302086.

\end